\documentclass[twocolumn,prd,amsmath,amssymb,superscriptaddress,nofootinbib]{revtex4}
\pdfoutput=1

\usepackage{epsfig,psfrag,graphics,verbatim}
\usepackage{graphicx}
\usepackage{dcolumn}
\usepackage{bm}
\usepackage{xcolor}
\usepackage{float}

\begin{document}

\title{Testing MOG theory in the Milky Way} 

\author{Carolina Negrelli}
\affiliation{Grupo de Astrof\'{\i}sica, Relatividad y Cosmolog\'{\i}a, 
        Facultad de Ciencias Astron\'{o}micas y Geof\'{\i}sicas, 
        Universidad Nacional de La Plata,
        Paseo del Bosque S/N 1900 La Plata, 
        Argentina}
\affiliation{CONICET, Godoy Cruz 2290, 1425 Ciudad Autónoma de Buenos Aires, Argentina}        

\author{Maria Benito}
\affiliation{ICTP--South American Institute for Fundamental Research, and Instituto de F\'isica Te\'orica / Universidade Estadual Paulista (UNESP), Rua Dr. Bento Teobaldo Ferraz 271, 01140-070 S\~{a}o Paulo, SP Brazil}

\author{Susana Landau}
\affiliation{Departamento de F\'{\i}sica and IFIBA, 
        Facultad de Ciencias Exactas y Naturales, 
        Universidad de Buenos Aires , 
        Ciudad Universitaria - Pab. I, 
        Buenos Aires 1428, 
        Argentina}
\affiliation{CONICET, Godoy Cruz 2290, 1425 Ciudad Autónoma de Buenos Aires, Argentina}  

\author{Fabio Iocco}
\affiliation{ICTP--South American Institute for Fundamental Research, and Instituto de F\'isica Te\'orica / Universidade Estadual Paulista (UNESP), Rua Dr. Bento Teobaldo Ferraz 271, 01140-070 S\~{a}o Paulo, SP Brazil}

\author{Lucila Kraiselburd}
\affiliation{Grupo de Astrof\'{\i}sica, Relatividad y Cosmolog\'{\i}a, 
        Facultad de Ciencias Astron\'{o}micas y Geof\'{\i}sicas, 
        Universidad Nacional de La Plata,
        Paseo del Bosque S/N 1900 La Plata, 
        Argentina}
\affiliation{CONICET, Godoy Cruz 2290, 1425 Ciudad Autónoma de Buenos Aires, Argentina}

\date{\today}
\begin{abstract}
We perform a test of John Moffat's Modified Gravity theory (MOG) within the Milky Way, adopting the well known ``Rotation Curve'' method. We use the dynamics of observed tracers within the disk to determine the gravitational potential as a function of galactocentric distance, and compare that with the potential that is expected to be generated by the visible component only (stars and gas) under different ``flavors'' of the MOG theory, making use of a state--of--the--art setup for both the observed tracers and baryonic morphology.
Our analysis shows that in both the original and the modified version (considering a  self-consistent evaluation of the Milky Way mass), the theory fails to reproduce the observed rotation curve. We conclude that in none of its present formulation, the MOG theory is able to explain the observed Rotation Curve of the Milky Way.

\end{abstract}

\maketitle

\section*{Introduction}

A {\it dark} component of matter has become one of the pillars of current $\Lambda$CDM model: it is invoked to explain the mismatch between the observed dynamical mass, and that inferred by observations of the visible component, of astrophysical objects over a large range of mass and spatial scales, from Galaxy Clusters \cite{1998ApJ...498L.107T,2013JCAP...07..008H,2018MNRAS.475..532O,2016A&A...594A..24P} to Spiral \cite{2003A&A...397..899S,2008ApJ...684.1143X,2014ApJ...794...59K} and Dwarf Galaxies \cite{2017PhRvD..95h2001A}, including our own, \cite{2015NatPh..11..245I}, and provides a consistent explanation to the power spectrum of the Cosmic Microwave Background \cite{2016A&A...594A..13P}, and to the formation of astrophysical structures  \cite{2013MNRAS.432..743N}. Yet, the very nature of this {\it dark matter} is currently unknown, and none of the proposed candidates (from stable particles in extensions of the Standard Model, to primordial Black Holes \cite{2015PrPNP..85....1K,2016PhRvL.116t1301B}) has been unambiguously detected yet.

An alternative proposal to explain the mismatch observed in the data relies on a modification of the theory of gravity.  Several proposals, such as MOND, TeVeS and MOG \cite{1983ApJ...270..365M,2004PhRvD..70h3509B,2006JCAP...03..004M}, have been able to give an explanation to  phenomena around data coming from numerous and diverse sources:  motion of globular and galaxy clusters \cite{2008ApJ...680.1158M, 2014MNRAS.441.3724M, 2017EPJP..132..417M} and rotation curves of spiral and dwarf galaxies \cite{2013MNRAS.436.1439M,2017MNRAS.468.4048Z}. 

While some analysis indicate that TeVeS and MOG have difficulties explaining the Bullet cluster data \cite{2006ApJ...648L.109C} or to reconcile gas profile and strong--lensing measurements in well known cluster systems \cite{2018MNRAS.tmp..370N}, others claim that MOG can fit both Bullet and the Train Wreck merging clusters \cite{2007MNRAS.382...29B,2016arXiv160609128I}.
It has been pointed out that the detection of a neutron star merger by the LIGO experiment rules out MOND-like theories \cite{2017arXiv171006168B}.  Recent analysis state 
the former being correct for bi-metric theories such as MOND and TeVeS, but not for MOG \cite{2017arXiv171011177G}. Some of the above controversies are yet to be resolved, so it is currently unclear if MOG phenomenology can offer a solution at all scales.

In this work, we adopt an agnostic approach, and only focus on the prediction of MOG theory on the scale of Spiral Galaxies, with a specific one: our own host. In order to test the predictions of MOG theory within the Milky Way, we use state--of--the--art compilations of kinematical tracers and observationally inferred morphologies, adopted in recent studies of Dark Matter distribution \cite{Iocco:2011jz, Pato:2015dua, 2015NatPh..11..245I}, and already used to test MOND phenomenology \cite{Iocco:2015iia}.

\section{MOG theories}
\label{theory}
 The theory includes  a massive vector field $\phi^\mu$ and three scalar fields: $G$, $\mu$, $\omega$  which represent the gravitational coupling strength, the mass of the vector field and its coupling strength respectively. The last one is a dimensionless field commonly taken  as 1. The gravitational action can be expressed as:
\begin{equation}
S_G=-\frac{1}{16\pi}\int\frac{1}{G}\left({\it R}+2\Lambda\right)\sqrt{-g}~d^4x,
\end{equation}
Besides,  the massive vector field $\phi_\mu$ action is:
\begin{eqnarray}
S_\phi&=&-\frac{1}{4\pi}\int\omega\Big[\frac{1}{4}{\bf\it B^{\mu\nu}B_{\mu\nu}}-\frac{1}{2}\mu^2\phi_\mu\phi^\mu\nonumber\\
&+&V_\phi(\phi_\mu\phi^\mu)\Big]\sqrt{-g}~d^4x,
\end{eqnarray}
where $B_{\mu\nu}=\partial_\mu\phi_\nu-\partial_\nu\phi_\mu$ 
is the Faraday tensor of the vector field. 
The scalar fields action can be written as: 
\begin{eqnarray}
S_S=-&\int\frac{1}{G}\Big[\frac{1}{2}g^{\alpha\beta}\biggl(\frac{\nabla_\alpha G\nabla_\beta G}{G^2}
+\frac{\nabla_\alpha\mu\nabla_\beta\mu}{\mu^2}\biggr)\nonumber\\
&+\frac{V_G(G)}{G^2}+\frac{V_\mu(\mu)}{\mu^2}\Big]\sqrt{-g}~d^4x.
\label{scalar}
\end{eqnarray}
Here, $\nabla_\nu$ is the covariant derivative with respect to
the metric $g_{\mu\nu}$ and  $V_\phi(\phi_\mu\phi^\mu)$, $V_G(G)$ and $V_\mu(\mu)$ are the self-interaction potentials associated with the
vector field and the scalar fields, respectively. 

For studying the behavior of MOG on astrophysical scales we can use the weak field approximation for the dynamics of the fields, perturbing them around Minkowski space time for an arbitrary distribution of non relativistic matter. Under this scheme, the scalar fields remain constant. Following \cite{2013MNRAS.436.1439M} the 
acceleration of a test particle as the gradient of the effective potential $(\vec{a} = -\vec{\nabla}\Phi_{\rm eff})$ can be written as,

\begin{align}
\vec{a}(\vec x) =& - G_N\int\frac{\rho(\vec x')(\vec{x}-\vec{x'})}{|\vec x-\vec x'|^3}\nonumber\\
&{}\times\left[1+\alpha -\alpha e^{-\mu|\vec x-\vec x'|}(1+\mu|\vec x-\vec x'|) \right]d^3\vec{x}'. \label{acceleration}
\end{align}
 The parameter $\alpha$ and the vector field mass  $\mu$ control the strength  and the range of the ``fifth force'' interaction respectively, and their estimates made in \cite{2009CQGra..26h5002M} as functions of the mass are given by\footnote{This expression  is derived for a spherically symmetric point-like source which is not the situation at hand. However we consider this analytical expression as the first guess for $\alpha$ and $\mu$.},

\begin{equation}
\alpha=\frac{M}{(\sqrt{M}+E)^2}\left(\frac{G_\infty}{G_N}-1\right),
\label{eq:alpha2}
\end{equation}
and
\begin{equation}
\mu=\frac{D}{\sqrt{M}},
\label{eq:mu}
\end{equation}
where $\mu$ is in units of kpc$^{-1}$. Hereafter, the dimension of $\mu$ will not be specified anymore.
$G_{\infty}\simeq 20 G_{N}$ represents the effective gravitational constant at infinity, while $D$ and $E$ are determined using observational data \cite{2009CQGra..26h5002M}.

\section{Methodology and setup}
\label{method}
In order to test the most common MOG scenarios with our Galaxy, we use a comprehensive compilation of kinematic tracers of the Milky Way and a state-of-the-art modeling of the baryons, both presented in Ref.~\cite{2015NatPh..11..245I}.
We improve the analysis over previous ones in the Milky Way,  \cite{2015PhRvD..91d3004M}, by:
{\it a)} adopting --separately-- two compilations  tracers of the rotation curve, that have a higher density of data in the galactocentric distances $2.5< R < 25$ {\rm kpc}, whereas the dataset adopted in \cite{2015PhRvD..91d3004M}, is denser in the interval $20<R<100$ {\rm kpc};
{\it b)} for the rotation curve expected by the baryonic component, implementing a full set of three-dimensional observationally-inferred baryonic morphologies including bulge, disk, and gas component, and solve the integral in Eq. \ref{acceleration} numerically, whereas  \cite{2015PhRvD..91d3004M} employs only an analytical formulation \cite{2012PASJ...64...75S}.

As observed tracers of the gravitational potential, we adopt the compilation of halo star data from \cite{huang2016milky} (hereafter ``Huang''), which extends up to 100 kpc. We also test our final results against the compilation of tracers {\tt galkin} first presented in \cite{2015NatPh..11..245I} (and Supplementary Information therein) and then publicly released in \cite{Pato:2017yai}, that offers an enhanced number of diverse types of objects within the disk, in innermost regions of the Milky Way.
Our conclusions remain qualitatively unchanged when using the two compilations, based on different types of objects, subject to different analysis, and in different regions of the MW. 

To model the density field of the baryonic content (stars and gas) within the Milky Way, we adopt a set of observationally inferred morphologies, separating the stellar component in bulge and disk, and also accounting for the interstellar gas. 
By combining individually a selected choice of each component (and then varying one at the time) we obtain a large array of individual morphologies which bracket the systematic uncertainty on the distribution of the baryonic mass our Galaxy.

We follow the technique first presented in \cite{Iocco:2011jz}, 
summarizing here its most crucial points, we address to the original publications \cite{Iocco:2011jz, 2015NatPh..11..245I, Pato:2015dua} for further details.

Bulge and disk density profiles are individually normalized to the MACHO microlensing optical depth observation in the Galactic Center region \cite{Popowski:2004uv}, 
and to the surface stellar density measurement \cite{Bovy:2013raa}, respectively. Both observations carry statistical uncertainties propagating to the normalization of the bulge and disk.

Together with the gas component, a statistical uncertainty is thus associated to the total baryonic density of the Galaxy. 
This propagates to the rotational velocity computed through the gravitational potential, allowing a statistically meaningful test of the rotation curve obtained for each single morphology.

We integrate these full three-dimensional density functions of bulge, disk and gas through equation \ref{acceleration} 
in order to obtain the MOG acceleration at each galactocentric distance, and its corresponding circular velocity at the Galactic plane, i.e. z=0.

The rotation curve for the baryonic component under the MOG potential is compared to the observed rotation curve, building a $\chi^2$ for the angular velocities $w(R)=v_c(R)/R$, adopting the uncertainties on the observed RC, and that for baryons as described above;  for both compilations the data are taken individually, without binning.
When the {\tt galkin} compilation is adopted, instead of the usual definition of $\chi^2$
we use the function described in \cite{2015NatPh..11..245I} (Eq. 2 of Supplementary Information), which account for uncertainties both in angular velocities and galactocentric distances, and it has been shown to have a $\chi^2$ distribution
\footnote{We only include data points at $R>$ $R_{cut}=$2.5 kpc in the analysis, in order to avoid spurious results due to departure from cylindrical symmetry of the galactic bulge, e.g. \cite{Catena:2009mf, 2015NatPh..11..245I}.  Further tests of the validity of the results against the departure from circularity and most relevant sources of asymmetry are performed in the Supplementary Information of \cite{2015NatPh..11..245I}.
We adopt $\rm (R_0, V_0)=(8.35\,\,kpc, 239.89\,\,km/s)$ and $\rm (U, V, W)_{\odot} = (7.01, 10.13, 4.95)\,\,km/s$. Varying these Galactic parameters within the currently accepted range of systematic uncertainties does not modify our conclusions, \cite{2015NatPh..11..245I, Iocco:2015iia}.}.

\section{Results}
\label{results}
We use the setup built above in order to test the MOG theory for each single morphology in our catalogue. For the sake of clarity, we first describe results for a single morphology, denoted a ``{\it representative}'', composed of the disk in \cite{CalchiNovatiMancini2011}, the {\it E2} bulge in \cite{Dwek1995}, and the gas \cite{Ferriere1998,Ferriere2007} (number 8 in the Table \ref{tab:chi2}). We generalize our results to all other morphologies at the end of this Section.

 We test the MOG theory in its ``standard'' formulation adopting the couple of parameters ($\alpha,\mu$) indicated by Moffat  as the best possible values to fit Spiral Galaxies \cite{2013MNRAS.436.1439M} ($\alpha,\mu$)$^{\rm SG}$ and Milky Way \cite{2015PhRvD..91d3004M} ($\alpha,\mu$)$^{\rm MW}$. The latter are obtained through Eq. \ref{eq:alpha2} and \ref{eq:mu} with a Milky Way baryonic mass of $\rm M^{MW}_{Mof}$=$4\times 10^{10}\,{\rm M_{\odot}}$. Parameters ($\alpha,\mu$)$^{\rm C}$ are set using the same approach but a with different mass value $\rm M^{MW}_C$=$(6.7^{+0.7}_{-0.6})\times 10^{10}\,{\rm M_{\odot}}$ that we self-consistently obtain from our morphological model.


In Table \ref{tab:chi2}, row \# 8, we show the values of reduced $\chi^2$ for each of these three set of parameters, which falls beyond the 5 $\sigma$ equivalent $\tilde{\chi}^2_{5\sigma}$ (2.41  for Huang --43 d.o.f-- and 1.14 for {\tt galkin} --2701 d.o.f--), thus indicating that for this morphology, MOG theory with these parameters is ruled out with a large degree of confidence.

 A check by using the recent-most data compilation {\tt galkin}
finds no qualitative change, leaving intact the above conclusion. The difference in absolute values of ${\chi}^2$ between the two compilations reflects the sensitivity of the two datasets to different regions of the MW.

\begin{figure*}[t]
\centering
\includegraphics[width=0.9\textwidth]{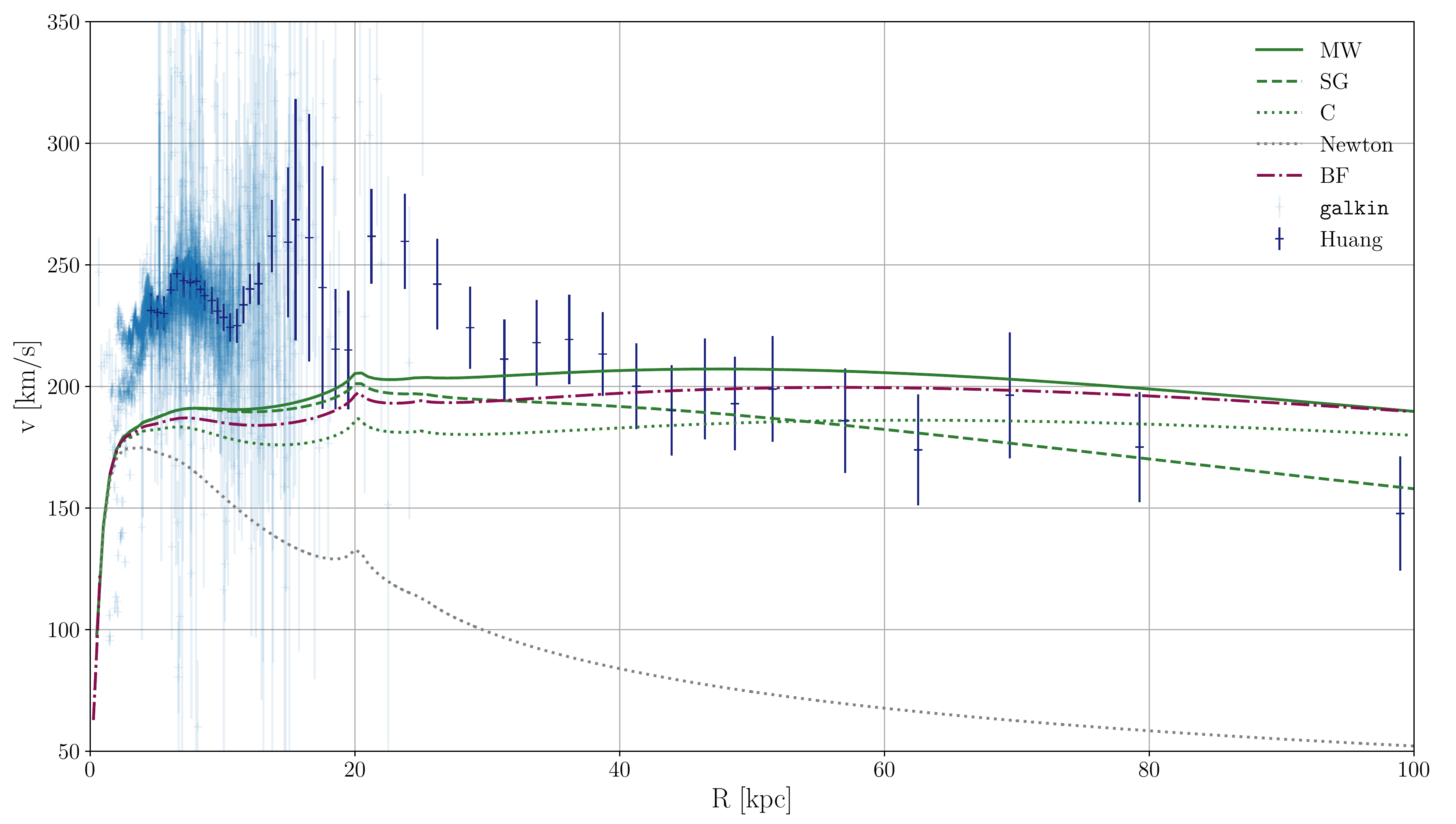}
\caption{Rotation curves for the MOG ``flavour'': relevant cases MW, SG, C, BF (see text for definition and details), for our representative morphology.
Uncertainties are shown only for the observational data, with the central value only displayed for ``MOG--expected'' baryonic rotation curves, to ease visualization.}
\label{fig:rotcurve_representative}  
\end{figure*}

\par  Existing work, \cite{2009PhDT.......172B}, assigns an uncertainty to $D = (6.44 \pm 0.20)\,\rm M_{\odot}^{1/2}{pc}^{-1}$ and $E=(28.4 \pm 7.9)\times 10^3\,\rm M_{\odot}^{1/2}$, which propagates to the values of ($\alpha,\mu$)
when applying Eqs. \ref{eq:alpha2} and \ref{eq:mu} to the value of the baryonic mass of the Galaxy with its uncertainties.

We thus obtain the parameter interval $\alpha=15.4\pm 1.0$ and $\mu=(2.5\pm0.2)\times10^{-2}$.
We scan this interval, and find that for each point in this two-dimensional space, the reduced $\chi^2$ is beyond the 5 $\sigma$ equivalent (we use the Huang compilation's $\tilde{\chi}^2$ as reference in the scan), with the lowest one being $\tilde{\chi}^2_{BF}=8.60$, for the parameter point ($\alpha,\mu$)$_{\rm BF}=(16.4, 2.7\times10^{-2})$.
This bears the  conclusion that MOG theory fails to explain the observed rotation curve of the Milky Way, for the morphology at study. 


As appreciable from both Table \ref{tab:chi2} row \# 8, and Figure \ref{fig:rotcurve_representative}, MOG admittedly performs better than Newtonian gravity, but fails to describe the shape of the observed Rotation Curve.

\par We now extend our methodology to the entire set of morphologies contemplated by previous studies. 
It is worth to recall here that each possible morphology is alternative to another one, and it is not possible to infer from their ensemble any median, mean, or ``typical'' value. However, they represent a nearly complete set of all possible morphologies still considered viable to date, and their spread can be considered a satisfactory indicator of the systematics present for the Milky Way, with the conclusion that the actual physical reality must reasonably lie within them.

\par Separately and for each of the morphologies, we self-consistently compute the baryonic mass, and identify the corresponding ``corrected'' point in the $(\alpha, \mu)$ by applying Eqs. \ref{eq:alpha2} and \ref{eq:mu}.
We then produce the rotation curve and its uncertainties, and compute the reduced $\chi^2$ by using the Huang data compilation.

\setlength{\tabcolsep}{2pt}
\def\arraystretch{1.6}
\begin{table*}[!htbp]
\begin{tabular}{ l | c | c | c | c | c | c }
\hline
baryonic  & Newton  & MW  & SG & C  & $(\alpha, \mu)^{\rm C}$ & ${\rm M_{C}^{MW}}$ $[10^{10} \; {\rm M_{\odot}}]$ \\  
 morphology &$\tilde{\chi}^2$ &$\tilde{\chi}^2$ &$\tilde{\chi}^2$ &$\tilde{\chi}^2$ & & \\ 
$[$disk$]$ [bulge] & Huang --  {\tt galkin} &Huang --  {\tt galkin}  &Huang --  {\tt galkin} & Huang  --  {\tt galkin}  & & \\ \hline
 1 \cite{HanGould2003}\cite{Dwek1995} {\it G2} & 31.83 -- 10.69 & 4.50 -- 4.25 & 4.68 -- 4.25 &  8.59 -- 5.96 & (15.79, $2.43\times 10^{-2}$) & $6.6^{+0.6}_{-0.4}$ \\ \hline
 2 \cite{HanGould2003}\cite{Dwek1995} {\it E2} & 30.80 -- 9.89 & 4.11 -- 3.83 & 4.25 -- 3.83 & 8.00 -- 5.39 & (15.80, $2.41\times 10^{-2}$) & $6.7^{+0.7}_{-0.6}$ \\ \hline
 3 \cite{HanGould2003}\cite{Vanhollebeke2009} & 32.90 -- 8.51 & 3.36 -- 3.10 & 3.43 -- 3.10 & 6.85 -- 4.37 & (15.83, $2.39\times 10^{-2}$)  & $6.8^{+0.7}_{-0.6}$ \\ \hline
 4 \cite{HanGould2003}\cite{BissantzGerhard2002} & 29.85 -- 9.45 & 3.71 -- 3.51 & 3.79 -- 3.51 & 7.47 -- 5.03 & (15.83, $2.39\times 10^{-2}$) & $6.8^{+0.7}_{-0.6}$  \\ \hline
 5 \cite{HanGould2003}\cite{Zhao1996} & 35.73 -- 11.40 & 4.93 -- 4.66 & 5.16 -- 4.66 & 9.21 -- 6.51 & (15.77, $2.44\times 10^{-2}$) & $6.6\pm0.6$ \\ \hline
 6 \cite{HanGould2003}\cite{Robin2012} & 28.67 -- 13.65 & 6.17 -- 6.00 & 6.48 -- 6.00 & 13.00 -- 8.43  & (15.74, $2.47\times 10^{-2}$)  & $6.4^{+0.6}_{-0.5}$ \\ \hline
 7 \cite{CalchiNovatiMancini2011}\cite{Dwek1995} {\it G2} & 33.84 -- 12.69 & 5.51 -- 5.45 & 5.74 -- 5.44 & 9.86 -- 7.37 & (15.79, $2.42\times 10^{-2}$) & $6.6^{+0.6}_{-0.4}$ \\ \hline
 8 \cite{CalchiNovatiMancini2011}\cite{Dwek1995}  {\it E2}  & 32.65 -- 11.72 & 5.02 -- 4.90 & 5.20 -- 4.90 & 9.14 -- 6.65 & (15.80, $2.41\times 10^{-2}$) & $6.7^{+0.7}_{-0.6}$ \\ \hline
 9 \cite{CalchiNovatiMancini2011}\cite{Vanhollebeke2009} & 30.19 -- 10.04 & 4.06 -- 3.93 & 4.17 -- 3.93 & 7.72 -- 5.23  & (15.84, $2.38\times 10^{-2}$) & $6.9^{+0.7}_{-0.6}$ \\ \hline
 10 \cite{CalchiNovatiMancini2011}\cite{BissantzGerhard2002} & 31.62 -- 11.22 & 4.54 -- 4.50 & 4.66 -- 4.50 & 8.53 -- 6.22 & (15.83, $2.39\times 10^{-2}$) & $6.9^{+0.7}_{-0.6}$  \\ \hline
 11 \cite{CalchiNovatiMancini2011}\cite{Zhao1996} & 35.10 -- 13.56 & 6.06 -- 5.98 & 6.33 -- 5.97 & 10.64 -- 8.10 & (15.77, $2.44\times 10^{-2}$) & $6.6\pm0.6$ \\ \hline
 12 \cite{CalchiNovatiMancini2011}\cite{Robin2012} & 38.46 -- 16.32 & 7.66 -- 7.74 & 8.03 -- 7.74 & 15.79 -- 10.60 & (15.73, $2.47\times 10^{-2}$) & $6.4^{+0.6}_{-0.5}$ \\ \hline
 13 \cite{deJong2010}\cite{Dwek1995} {\it G2} & 33.70 -- 12.39 & 5.43 -- 5.29 & 5.66 -- 5.28 & 9.80 -- 7.17 & (15.79, $2.42\times 10^{-2}$) & $6.7^{+0.6}_{-0.4}$ \\ \hline
 14 \cite{deJong2010}\cite{Dwek1995} {\it E2} & 32.54 -- 11.45 & 4.94 -- 4.76 & 5.15 -- 4.76 &  9.09 -- 6.47 & (15.81, $2.41\times 10^{-2}$) & $6.7^{+0.7}_{-0.6}$ \\ \hline
 15 \cite{deJong2010}\cite{Vanhollebeke2009} & 30.14 -- 9.82 & 4.02 -- 3.83 & 4.14 -- 3.83 & 7.71 -- 5.11 & (15.84, $2.38\times 10^{-2}$) & $6.9^{+0.7}_{-0.6}$ \\ \hline
16 \cite{deJong2010}\cite{BissantzGerhard2002} & 31.50 -- 10.95 & 4.46 -- 4.37 & 4.60 -- 4.37 & 8.49 -- 6.06 & (15.84, $2.38\times 10^{-2}$)  & $6.9^{+0.7}_{-0.6}$ \\ \hline
17 \cite{deJong2010}\cite{Zhao1996} & 34.93 -- 13.23 & 5.96 -- 5.80 & 6.24 -- 5.79 & 10.56 -- 7.86 &  (15.78, $2.44\times 10^{-2}$)  & $6.6\pm0.6$  \\ \hline
18 \cite{deJong2010}\cite{Robin2012} & 38.18 -- 15.89 & 7.5 -- 7.48 & 7.87 -- 7.47 & 15.49 -- 10.27 & (15.74, $2.47\times 10^{-2}$) & $6.4^{+0.6}_{-0.5}$ \\ \hline
19 \cite{Juric2008}\cite{Dwek1995} {\it G2} & 32.81 -- 11.45 & 5.22 -- 4.91 & 5.18 -- 4.90 & 8.46 -- 5.96 & (15.91, $2.32\times 10^{-2}$) & $7.2^{+0.6}_{-0.5}$ \\ \hline
20 \cite{Juric2008}\cite{Dwek1995} {\it E2} & 31.79 -- 10.66 & 4.79 -- 4.48  & 4.76 -- 4.47 & 7.86 -- 5.35 & (15.92, $2.31\times 10^{-2}$) & $7.3^{+0.7}_{-0.6}$ \\ \hline 
21 \cite{Juric2008}\cite{Vanhollebeke2009} & 33.86 -- 9.26 & 3.99 -- 3.71 & 3.99 -- 3.70 & 6.69 -- 4.35 & (15.95, $2.29\times 10^{-2}$)  & $7.5^{+0.8}_{-0.7}$\\ \hline
22 \cite{Juric2008}\cite{BissantzGerhard2002} & 30.64 -- 10.19 & 4.21 -- 4.07 & 4.20 -- 4.06 & 7.30 -- 5.01 & (15.95, $2.29\times 10^{-2}$) & $7.5^{+0.7}_{-0.6}$  \\ \hline
23 \cite{Juric2008}\cite{Zhao1996} & 36.51 -- 12.17 & 5.68 -- 5.33 & 5.63 -- 5.32 &  9.11 -- 6.45 & (15.89, $2.34\times 10^{-2}$) & $7.2^{+0.7}_{-0.6}$ \\ \hline
24 \cite{Juric2008}\cite{Robin2012} & 29.76 -- 14.42 & 6.91 -- 6.67 & 6.83 -- 6.66 & 12.91 -- 8.34 & (15.85, $2.37\times 10^{-2}$) &  $7.0\pm0.6$ \\ \hline
 25 \cite{Bovy:2013raa}\cite{Dwek1995} {\it G2} & 24.48 -- 4.87 & 1.94 -- 1.50 & 1.79 -- 1.51 &  4.50 -- 2.07  & (15.94, $2.30\times 10^{-2}$) & $7.4^{+0.7}_{-0.6}$ \\ \hline
 26 \cite{Bovy:2013raa}\cite{Dwek1995} {\it E2} & 24.02 -- 4.64 & 1.84 -- 1.42 & 1.68 -- 1.43 & 4.30 -- 1.97 & (15.94, $2.29\times 10^{-2}$) & $7.4^{+0.8}_{-0.7}$ \\ \hline
 27 \cite{Bovy:2013raa}\cite{Vanhollebeke2009} & 23.23 -- 4.15 & 1.70 -- 1.29 & 1.53 -- 1.29 & 3.97 -- 1.72 & (15.95, $2.29\times 10^{-2}$) & $7.5^{+0.8}_{-0.7}$ \\ \hline
 28 \cite{Bovy:2013raa}\cite{BissantzGerhard2002} &  22.9 -- 4.47 & 1.58 -- 1.26 & 1.32 -- 1.27 & 3.82 -- 1.84 & (15.98, $2.26\times 10^{-2}$)  & $7.7^{+0.8}_{-0.7}$ \\ \hline
 29 \cite{Bovy:2013raa}\cite{Zhao1996} & 24.93 -- 3.89 & 2.03 -- 1.58 & 1.90 -- 1.59 & 4.76 -- 2.20 & (15.93, $2.30\times 10^{-2}$)  & $7.4\pm0.7$ \\ \hline
 30 \cite{Bovy:2013raa}\cite{Robin2012} & 25.78 -- 5.88 & 2.20 -- 1.81 & 2.08 -- 1.81 & 5.40 -- 2.62 & (15.92, $2.31\times 10^{-2}$) &  $7.3^{+0.7}_{-0.6}$ \\ \hline
\end{tabular}
\caption{$\tilde{\rm \chi^2}$ for all MOG ``flavours'' (MW, SG, C), parameters obtained as described in text.
For all morphologies, gas profiles taken from \cite{Ferriere1998,Ferriere2007} are added to the quoted disk and bulge ({\rm G2} or {\rm E2} refers to different configurations in \cite{Dwek1995}), and MW baryonic mass computed self--consistently.}
\label{tab:chi2}
\end{table*}

\begin{figure*}[t]
\centering
\includegraphics[width=0.9\textwidth]{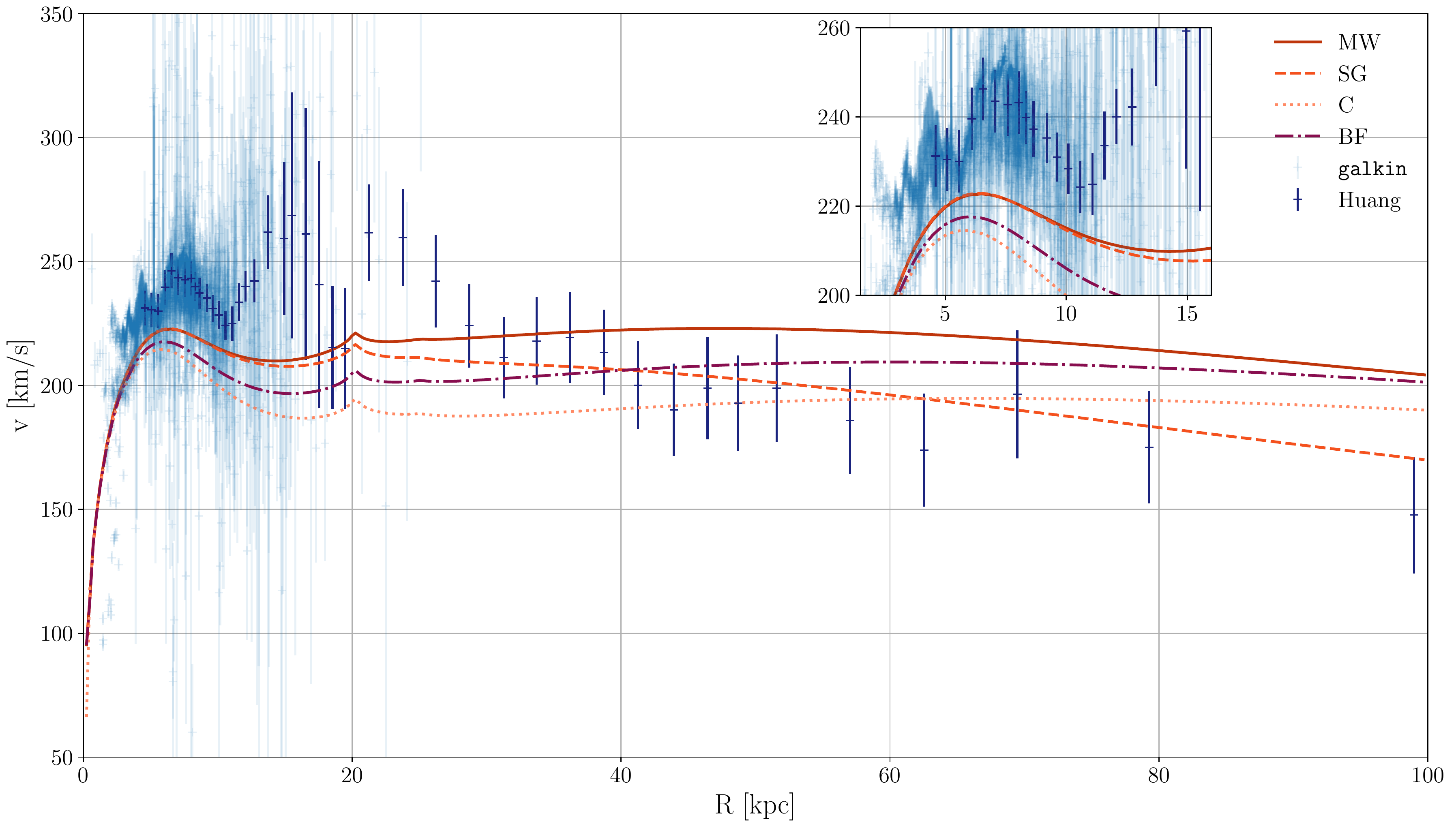}
\caption{Rotation curves for the MOG ``flavour'': relevant cases MW, SG, C, BF, for our ``best fitting'' morphology, \# 28.
Uncertainties are shown only for the observational data, with the central value only displayed for ``MOG--expected'' baryonic rotation curves, to ease visualization.}

\label{fig:rotcurve_BRBG-BF}  
\end{figure*}

\par  The reduced $\chi^2$ values for all morphologies are shown in table \ref{tab:chi2}, for parameters $(\alpha, \mu)^{\rm MW}$, $(\alpha, \mu)^{\rm SG}$ and $(\alpha, \mu)^{\rm C}$, respectively, showing disagreement between the MOG rotation curve and the observed one at more than 5 $\sigma$ for all morphologies with the exception of one set. Morphologies carrying the disk in \cite{Bovy:2013raa} (``BR disk'' hereafter), bear $\chi^2$ visibly better than others (while still excluded at more than 5 $\sigma$ equivalent when tested against the {\tt galkin} compilation) because this disk is heavier than the others considered, thus carrying the overall normalization of the obtained rotation curve closer to the observed one in the innermost regions.
We select the morphology that systematically produces the best $\chi^2$ (\# 28 in Table \ref{tab:chi2}), and we scan the parameter space around the point defined by the central value of the mass up to the point defined by the current uncertainty, as done for the representative morphology.
The baryonic mass for this morphology is ${\rm M_{C}^{MW}}=(7.7^{+0.8}_{-0.7})\times 10^{10}\,{\rm M_{\odot}}$, and the parameter space scanned
is $\alpha\in [14.7, 16.6]$ and $\mu \in [2.14, 2.52]\times 10^{-2}\,{\rm kpc^{-1}}$.
Within this range, the best fitting point is $(\alpha, \mu)^{\rm BF} = (16.6, 2.52\times10^{-2}\,{\rm kpc^{-1}})$, bearing the reduced $\tilde{\chi}^2$=2.78, which for the degrees of freedom of the Huang, is incompatible at more than 5 $\sigma$.

In Fig \ref{fig:rotcurve_BRBG-BF}, we show the data together with the rotation curve for this best--fitting morphology, for all the significant points (MW, SG, C) in the parameter space, including the best--fitting point.

None of these curves manages to capture the very behavior in the central 15 kpc --the entirety of the visible Milky Way-- always producing rotation curves below the observed ones.
We test the points above against the {\tt galkin} data compilation, richer of data in the region in object, and report the corresponding $\chi^2$ values in Table \ref{tab:chi2} and the above paragraph.
Whereas better than for any other case, they indicate an incompatibility at more than 5 $\sigma$ for all the cases in object, thus again bearing the conclusion that MOG theory can not explain the observed rotation curve of the Milky Way 
\footnote{We have tested our results against the inclusion of hot gas halo. Reduced $\chi^2$ remain above the five sigma threshold, and our conclusions unaffected.}.

\section{Conclusions}
\label{conclusions}
We have performed a test of MOG theories against the Milky Way dynamics, improving with respect to previous analysis
on one hand by using two recent--most compilation of data for the observed Rotation Curve, and on the other by adopting a virtually complete set of observationally inferred morphologies for the stellar and gas (baryonic) component. 
\par We have also modified the key--parameters of the theory, in order to match them to the baryonic mass of the Milky Way as self--consistently obtained within the morphologies we adopt, individually at each time. Once again, the obtained rotation curves disagree with the observed one with a strong statistical evidence for the entire set of morphologies.
\par In light of this analysis, we conclude that modifying the gravitational potential according to the current version of MOG theory, does not offer a viable solution to the discrepancy between the observed rotation curve, and that generated by the baryons only, in the Milky Way.

\vspace{0.5cm}
We thank J.~Moffat for valuable comments and suggestions on the analysis and the manuscript. F.~I.~would like to thank P.~D.~Serpico for the sharp eye, as usual.
C.~N., S.~L. and L.~K.  are supported by PIP 11220120100504 (CONICET) and grant G140 (UNLP). C.~N.~ and S.~L.~acknowledge hospitality in S\~ao Paulo through the ICTP-SAIFR FAPESP grant 2016/01343-7. F.~I.~acknowledges support from the Simons Foundation and FAPESP process 2014/11070-2.
This research was supported by resources supplied from the Center for Scientific Computing (NCC/GridUNESP) of the S\~ao Paulo State University (UNESP).

\bibliographystyle{apsrev}
\bibliography{testmog}

\end{document}